\documentclass[12pt]{article}

\newcommand{\pslash}{\not \! p}

\usepackage{amsmath,amssymb,amscd,amsbsy,amsgen,amsopn,amstext,
amsxtra}

\usepackage[dvips]{graphicx}

\begin{document}

\begin{center}
{\Large{\bf On neutrino masses via CPT violating Higgs interaction in the Standard Model}}
\end{center}
\vskip .5 truecm
\begin{center}
{\bf { Masud Chaichian{$^*$}, Kazuo Fujikawa$^\dagger$ and
Anca Tureanu$^*$}}
\end{center}

\begin{center}
\vspace*{0.4cm} {\it { $^*$Department of Physics, University of
Helsinki, P.O.Box 64, FIN-00014 Helsinki,
Finland\\
$^\dagger$ Mathematical Physics Laboratory, RIKEN Nishina Center,\\
Wako 351-0198, Japan}}
\end{center}


\begin{abstract}
The Lorentz invariant $CPT$ violation by using non-local interactions is
naturally incorporated in the Higgs coupling to neutrinos in the
Standard Model, without spoiling the basic $SU(2)_{L}\times U(1)$ gauge
symmetry. The neutrino--antineutrino mass splitting is thus realized
by the mechanism which was proposed recently, assuming the neutrino masses to be predominantly Dirac-type in the Standard Model.
\end{abstract}

\section{Introduction}
The $CPT$ symmetry is a fundamental symmetry of local field theory
defined in Minkowski space-time~\cite{pauli}. However, the possible
breaking of $CPT$ symmetry has also been discussed. One of the logical
ways to break $CPT$ symmetry is to make the theory non-local by
preserving Lorentz symmetry, while the other is to break Lorentz
symmetry itself. Lorentz symmetry breaking scheme has been mainly
discussed in the past~\cite{ellis}, but a possible mechanism to break
$CPT$ symmetry in a Lorentz invariant manner has also been
proposed~\cite{chaichian} (see also~\cite{JGB}). We then presented an
explicit non-local  Lagrangian model which induces the particle
antiparticle mass splitting in a Lorentz invariant
manner~\cite{chaichian2},
\begin{eqnarray}\label{(2.3)}
S&=&\int d^{4}x\Big\{\bar{\psi}(x)i\gamma^{\mu}\partial_{\mu}\psi(x)
  - m\bar{\psi}(x)\psi(x)\\
  && -\int
d^{4}y[\theta(x^{0}-y^{0})-\theta(y^{0}-x^{0})]\delta((x-y)^{2}-l^{2})\nonumber\\
  &&\times[i\mu\bar{\psi}(x)\psi(y)]\Big\},\nonumber
\end{eqnarray}
which is Lorentz invariant and hermitian. For the real parameter $\mu$,
the third term has $C=CP=CPT=-1$ and thus no symmetry to ensure the equality
of particle and antiparticle masses. The parameter $l$ has dimension
of length, and the mass dimension of the parameter $\mu$ is $[M]^{3}$.

The free equation of motion for the fermion in (1) is given by
\begin{eqnarray}\label{(2.4)}
&&i\gamma^{\mu}\partial_{\mu}\psi(x)=m\psi(x)\\
&&+i\mu\int
d^{4}y[\theta(x^{0}-y^{0})-\theta(y^{0}-x^{0})]\delta((x-y)^{2}-l^{2})\psi(y).\nonumber
\end{eqnarray}
By inserting an ansatz for the possible solution
$\psi(x)=e^{-ipx}U(p)$,
we have
\begin{eqnarray}\label{(2.6)}
\pslash U(p)&=&mU(p)
+i\mu[f_{+}(p)-f_{-}(p)]U(p),
\end{eqnarray}
where $f_{\pm}(p)$ is the Lorentz invariant "form factor" defined by
\begin{eqnarray}\label{(1.3)}
&&f_{\pm}(p)=\int d^{4}z_{1}
e^{\pm ipz_{1}}\theta(z_{1}^{0})\delta((z_{1})^{2}-l^{2}),
\end{eqnarray}
which are  inequivalent for the time-like $p$ due to the factor
$\theta(z_{1}^{0})$; this $f_{\pm}(p)$ is mathematically related to
the two-point Wightman function for a free scalar
field~\cite{chaichian2} and thus expected to be well-defined at least
as a distribution.
By assuming a time-like $p$, we go to the frame where $\vec{p}=0$.
Then the eigenvalue equation for the mass is given by
\begin{eqnarray}\label{(2.10)}
p_{0}&=&\gamma_{0}\left[m - 4\pi \mu\int_{0}^{\infty}dz\frac{z^{2}\sin [
p_{0}\sqrt{z^{2}+l^{2}}]}{\sqrt{z^{2}+l^{2}}}\right],
\end{eqnarray}
where we used  the explicit formula
\begin{eqnarray}\label{(1.4)}
f_{\pm}(p^{0})
&=&2\pi \int_{0}^{\infty}dz\frac{z^{2}e^{\pm
ip^{0}\sqrt{z^{2}+l^{2}}}}{\sqrt{z^{2}+l^{2}}}.
\end{eqnarray}
This eigenvalue equation under $p_{0}\rightarrow -p_{0}$ becomes
\begin{eqnarray}\label{(2.13)}
p_{0}
&=&\gamma_{0}\left[m + 4\pi\mu\int_{0}^{\infty}dz\frac{z^{2}\sin [
p_{0}\sqrt{z^{2}+l^{2}}]}{\sqrt{z^{2}+l^{2}}}\right],
\end{eqnarray}
which is not identical to the original equation in \eqref{(2.10)}.
This causes the mass splitting of particle and antiparticle in the
sense of Dirac, even if all $C$, $CP$ and $CPT$ symmetries are broken
in the present model.
See Ref.~\cite{chaichian2} for further details.

From the point of view of particle phenomenology, there is a strong
interest in the possible mass splitting between the neutrino and
associated antineutrino~\cite{murayama,altarelli, adamson}. The
purpose of the present letter is to discuss the application of the
above mass splitting mechanism to Dirac-type neutrinos in the Standard
Model.

\section{Beyond the Standard Model}
In the original Standard Model~\cite{weinberg}  the neutrinos are
assumed to be massless, but recent experiments indicate non-vanishing
neutrino masses. We thus go beyond the original Standard Model
by including massive neutrinos.

We study a one-generation model of leptons to explain the
essence of the mechanism.
We consider a minimal extension of the Standard
Model by
incorporating the right-handed neutrino:
\begin{equation}
  \psi_{L}=\left(
  \begin{array}{c}
   \nu_{L}\\ e_{L}
  \end{array}
  \right), \ \
  \psi_{R}=\left(
  \begin{array}{c}
   \nu_{R}\\ e_{R}
  \end{array}
  \right)
\end{equation}
and the part of the Standard Model Lagrangian relevant to our discussion
is given by
\begin{eqnarray}\label{standard}
{\cal L}&=&\overline{\psi}_{L}i\gamma^{\mu}
(\partial_{\mu} - igT^{a}W_{\mu}^{a}
             - i\frac{1}{2}g^{\prime}Y_{L}B_{\mu})\psi_{L}
\nonumber\\
         && +\overline{e}_{R}i\gamma^{\mu}(\partial_{\mu}
             + ig^{\prime}B_{\mu})e_{R}
+\overline{\nu}_{R}i\gamma^{\mu}\partial_{\mu}\nu_{R}\nonumber
\\
            &&-\left[
\frac{\sqrt{2}m_{e}}{v}\overline{e}_{R}\phi^{\dagger}\psi_{L}
+\frac{\sqrt{2}m_{D}}{v}\overline{\nu}_{R}\phi_{c}^{\dagger}\psi_{L}
             +\frac{m_{R}}{2}\nu_{R}^{T}C\nu_{R}\right]+ h.c.
\end{eqnarray}
with $Y_{L}=-1$, and the Higgs doublet and its $SU(2)$ conjugate:
\begin{eqnarray}
  \phi=\left(
  \begin{array}{c}
   \phi^{+}\\ \phi^{0}
  \end{array}
  \right), \ \ \ \ \ \
   \phi_{c}\equiv i\tau_{2}\phi^{\star}=\left(
  \begin{array}{c}
   \bar{\phi}^{0}\\ -\phi^{-}
  \end{array}\right).
\end{eqnarray}
The operator $C$ stands for the charge-conjugation matrix for
spinors. The term with $m_{R}$ in the above Lagrangian is the
Majorana mass term for the right-handed neutrino~\cite{seesaw}.

We take the above Lagrangian as a
{\em low-energy effective theory} and apply  to it the naturalness
argument of 't Hooft ~\cite{'t hooft}. We first argue that the
choice $m^{2}_{D}\gg m_{R}^{2}$ is natural, since by setting $m_{R}=0$
one recovers an enhanced fermion number symmetry in
\eqref{standard}~\cite{wolfenstein, langacker, extra-dim}. We then argue
that $m_{e}\gg m_{D}$ is also natural, since by setting $m_{D}=m_{R}=0$
one finds an enhanced symmetry
$\nu_{R}(x)\rightarrow \nu_{R}(x)+ \xi_{R}$, with constant $\xi_{R}$, in
the Lagrangian \eqref{standard}~\cite{fujikawa1}. Thus, our
basic assumption in the present letter is $m_{e}\gg m_{D}\gg m_{R}$,
namely, the so-called pseudo-Dirac scenario~\cite{wolfenstein}, and in
the explicit analysis below we adopt the Dirac limit $m_{R}=0$ for
simplicity.

Our next observation is that the combination
\begin{eqnarray}
\phi_{c}^{\dagger}(x)\psi_{L}(x)
\end{eqnarray}
is invariant under the full $SU(2)_{L}\times U(1)$ gauge symmetry. One
may thus add a hermitian non-local Higgs coupling, which is analogous
to the last term in (1), to the Lagrangian  \eqref{standard},
\begin{eqnarray}
{\cal L}_{CPT}(x)=-i\frac{2\sqrt{2}\mu}{v}\int
d^{4}y&\delta((x-y)^{2}-l^{2})\theta(x^{0}-y^{0})\big\{\bar{\nu}_{R}(x)\left(\phi_{c}^{\dagger}(y)\psi_{L}(y)\right)\nonumber\\
&-\left(\bar{\psi}_{L}(y)\phi_{c}(y)\right)\nu_{R}(x)\big\},
\end{eqnarray}
without spoiling the basic $SU(2)_{L}\times U(1)$ gauge symmetry.
In the unitary gauge, $\phi^{\pm}(x)=0$ and $\phi^{0}(x)\rightarrow (v
+ \varphi(x))/\sqrt{2}$, the neutrino mass  term (with $m_{R}=0$)
becomes
in terms of the action
\begin{eqnarray}\label{mass}
S_{\nu \rm mass}&=&\int
d^{4}x\Big\{-m_{D}\bar{\nu}(x)\nu(x)\left(1+\frac{\varphi(x)}{v}\right)\nonumber\\
&& -i\mu\int
d^{4}y\delta((x-y)^{2}-l^{2})\theta(x^{0}-y^{0})\nonumber\\
&&\times\left [\bar{\nu}(x)\left(1+\frac{\varphi(y)}{v}\right)(1-\gamma_{5})\nu(y)
-\bar{\nu}(y)\left(1+\frac{\varphi(y)}{v}\right)(1+\gamma_{5})\nu(x)\right]\Big\}
\nonumber\\
&=&\int d^{4}x\Big\{-m_{D}\bar{\nu}(x)\nu(x)\left(1+\frac{\varphi(x)}{v}\right)\nonumber\\
&& -i\mu\int
d^{4}y\delta((x-y)^{2}-l^{2})[\theta(x^{0}-y^{0})-\theta(y^{0}-x^{0})]
\bar{\nu}(x)\nu(y)\nonumber\\
&&+i\mu\int
d^{4}y\delta((x-y)^{2}-l^{2})\bar{\nu}(x)\gamma_{5}\nu(y)\nonumber
\\
&&-i\frac{\mu}{v}\int
d^{4}y\delta((x-y)^{2}-l^{2})\theta(x^{0}-y^{0})\nonumber\\
&&\times
[\bar{\nu}(x)(1-\gamma_{5})\nu(y)-\bar{\nu}(y)(1+\gamma_{5})\nu(x)]\varphi(y)\Big\}, 
\end{eqnarray}
where we have changed the naming of integration variables $x
\leftrightarrow y$ in some of the terms and used
$\theta(x^{0}-y^{0})+\theta(y^{0}-x^{0})=1$.

When one looks at the mass terms in \eqref{mass} without the Higgs
$\varphi$ coupling, the first two terms are identical to the two terms
in (1) but an extra parity-violating non-local mass term appears,
which adds an extra term $-i\mu\gamma_{5}g(p^{2})$ to $m$ in the mass
eigenvalue equations in (5) and (7); here $g(p^{2})=
\int d^{4}z_{1}e^{ipz_{1}}\delta((z_{1})^{2}-l^{2})$. This extra term
is $C$ and $CPT$ preserving and does not contribute to the mass
splitting.
  Since we are assuming that $CPT$ breaking terms are very small, we may
solve the mass eigenvalue equations iteratively by assuming that the
terms with
the parameter $\mu$ are much smaller than $m=m_{D}$. We then obtain
the mass eigenvalues of the neutrino and antineutrino at
\begin{eqnarray}
m_{\pm}
&\simeq&m_{D}-i\mu\gamma_{5} g(m_{D}^{2}) \pm
4\pi\mu\int_{0}^{\infty}dz\frac{z^{2}\sin [
m_{D}\sqrt{z^{2}+l^{2}}]}{\sqrt{z^{2}+l^{2}}},
\end{eqnarray}
where we have used the upper two (positive) components of the matrix
$\gamma_{0}$ in (5) and (7). The parity violating mass
$-i\mu\gamma_{5} g(m_{D}^{2})$ is now transformed away by a suitable
global chiral transformation without modifying the last term in (14)
to the order linear in the small parameter $\mu$. In this way, the
neutrino and antineutrino mass splitting is incorporated in the
Standard Model by a Lorentz invariant non-local $CPT$ breaking
mechanism, without spoiling the $SU(2)_{L}\times U(1)$ gauge symmetry. The
Higgs particle $\varphi$ itself has a tiny $C$-, $CP$- and $CPT$-violating coupling in (13).

\section{Discussion}

We have assumed Dirac-type neutrinos, but this may not be unnatural in the
present context since the notion of antiparticle is best defined
for a Dirac particle. In other words, if the neutrino--antineutrino mass splitting is confirmed by experiments, it would
imply that neutrinos are Dirac-type particles rather than
Majorana-type particles.
  Also, our identification of the neutrino mass terms as the origin of
the possible $CPT$ breaking may be natural if one recalls that the mass
terms of the neutrinos are
  the known origin of new physics beyond the original Standard Model.
The remaining couplings of the Standard Model are very tightly
controlled by the $SU(2)_{L}\times U(1)$ gauge symmetry, and one can
confirm that only the neutrino mass terms allow the present non-local
gauge invariant couplings without introducing Wilson-line type gauge
interactions. (An analysis of the scheme with Wilson-lines, which goes beyond the conventional local gauge principle, will be given elesewhere~\cite{CFT}.)

To apply our scheme to the analysis of neutrino phenomenology
including neutrino
oscillation, we need to generalize the scheme to the three generations of
neutrinos. We consider that the generalization including the neutrino
mixing does not present a difficulty of basic principle, although a
detailed analysis of the three generations of neutrinos and the
possible choice of the parameters $l$ and $\mu$
in our scheme is required. It could be that our scheme needs to be
generalized by introducing more free parameters to apply it to
realistic  particle phenomenology. Thus, our model may provide
an indirect support for the speculation on the possible mass splitting
between the neutrino and antineutrino~\cite{murayama}.

If such a splitting will indeed be observed by future experiments, the presented pseudo-Dirac scheme could be considered as an economical alternative to seesaw mechanism ~\cite{seesaw}, where at the same time an explanation for the mass splitting between the particle and its antiparticle is provided.


Finally, we would like to discuss some basic field theoretical issues
related to the non-local couplings in our scheme. As for the
quantization of the theory non-local in time, for which the notion of
canonical momentum is ill-defined, our suggestion is to use the path
integral on the basis of Schwinger's action principle. This path
integral is based on the equation of motion and provides correlation
functions which agree with the ordinary quantum mechanical
correlations for local theory; the canonical structure is recovered
later by means of Bjorken--Johnson--Low prescription~\cite{fujikawa}.
For non-local theory, this scheme provides a possible generalization
and provides a convenient scheme for the treatment of non-local terms
as small perturbation. 

It is also well-known that a theory non-local in
time generally spoils unitarity. In our scheme we treat  small
non-local couplings in (13) in the lowest order of perturbation, for which the
effects of the violation of unitarity are expected to be minimal. However, we have the neutrino propagator 
\begin{eqnarray}
\langle T^{\star}\nu(x)\bar{\nu}(y)\rangle
=\int \frac{d^{4}p}{(2\pi)^{4}} e^{-ip(x-y)}\frac{i}{\pslash-m_{D} +i\epsilon+i\mu\gamma_{5}g(p^{2})-i\mu[f_{+}(p)-f_{-}(p)]},
\end{eqnarray}
which includes the effects of non-local terms. In the pole approximation this propagator gives a sensible result in (14), but it may lead to difficulties in the off-shell domain. Alternatively, the $CPT$-violating terms in the presented scheme
as such could be regarded as the low-energy limit of a more basic theory
or coming from some higher-dimensional theories~\cite{extra-dim}, whose compactification would lead to non-local interactions, and thus the unitarity issue may be postponed to future study.
Otherwise, it is very gratifying that the basic $SU(2)_{L}\times U(1)$ gauge
symmetry together with Lorentz symmetry are exactly preserved by our non-local
$CPT$ violation. We can thus avoid the appearance of negative norm in
the gauge sector if one applies gauge invariant and Lorentz invariant
regularization.

\section*{Acknowledgements}

The support of the Academy of Finland under the Projects No. 136539 and 140886 is gratefully acknowledged.

\end{document}